\documentclass{PoS}
\usepackage{amsmath}

\title{Fisher zeros and RG flows for $SU(3)$ with $N_f$ flavors }

\ShortTitle{Fisher zeros and RG flows}

\author{Zechariah Gelzer\\
Department of Physics and Astronomy, University of Iowa, Iowa City, Iowa, USA\\
E-mail: \email{zechariah-gelzer@uiowa.edu}}

\author{\speaker{Yuzhi Liu}\\
Department of Physics and Astronomy, University of Iowa, Iowa City, Iowa, USA\\
Fermi National Accelerator Laboratory, Batavia, Illinois, USA\\
Department of Physics, University of Colorado, Boulder, Colorado, USA\\
E-mail: \email{yuzhi.liu@colorado.edu}}

\author{Yannick Meurice\\
Department of Physics and Astronomy, University of Iowa, Iowa City, Iowa, USA\\
E-mail: \email{yannick-meurice@uiowa.edu}}

\author{Donald Sinclair\\
HEP Division, Argonne National Laboratory\\
Department of Physics and Astronomy, University of Iowa, Iowa City, Iowa, USA\\
E-mail: \email{dks@hep.anl.gov}}

\abstract{We calculate the Fisher zeros for $SU(3)$ gauge theory with different $N_f$ flavors of staggered fermions for various values of the fermion mass. We discuss the finite-size scaling near the end point of the line of discontinuity of $\bar{\psi} \psi$ in the beta-mass plane and in the larger beta-lower mass region. We discuss possible interpretations of these results in terms of Wilsonian RG flows and their possible relevance to construct composite Higgs models.}

\FullConference{31st International Symposium on Lattice Field Theory - LATTICE 2013\\
		July 29 - August 3, 2013\\
		Mainz, Germany}

\begin{document}

\section{Motivation}

The  discovery of a ``Higgs-like'' particle with mass of $126$~GeV puts severe restrictions on possible beyond the standard model scenarios where this particle appears as a strongly coupled meson-like state. A lot of efforts have been put to exploring gauge theories with an infrared fixed point (IRFP) \cite{Neil:2012cb,Giedt:2012it,kuti2013} . Despite significant computational efforts, it is still quite difficult to figure out what kind of new continuum limits can be taken in the vicinity of  hypothetical IRFPs. These calculations require large volumes, small fermion masses and the ability to deal effectively with irrelevant directions. 
Before deciding if such theories can provide phenomenologically viable alternatives to the standard Brout-Englert-Higgs mechanism, it is important to understand their global renormalization group (RG) flows. In these proceedings, we attempt to provide a qualitative description of these flows and how they can be studied more quantitatively using the finite-size scaling of the Fisher zeros. 

In practice, the systems under study are always finite in the numerical simulations and the correlation length $\xi$ is bonded by the actual size of the system simulated and thus never diverges. Finite-size scaling hypothesis provides a tool to extract infinite volume system information from a finite system. Lee and Yang \cite{Lee-Yang1} were able to relate the phase transition, or the divergence of $\xi$,  to the pinching of the partition function zeros in the complex $z = e^{2\beta h}$ plane. Here $\beta$ is the inverse temperature and $h$ is the external magnetic field. It was later observed by Fisher \cite{Fisher} that the pattern of the partition function zeros in the complex temperature plane also gives information about the nature of the phase transition. We usually call the partition function zeros in the complex $z = e^{2\beta h}$ plane Lee-Yang zeros \cite{Ejiri:2005ts,Nakamura:2013ska,Ejiri:2013swa} and those in the complex temperature plane Fisher zeros. Interestingly, one can perform finite-size scaling analysis on the zeros and extract critical exponents from them \cite{Liu:2011zzh,  Meurice:2012sj, Denbleyker:2013bea}.
It was also argued that Fisher zeros act as separatrices for the RG flows in the complex coupling plane \cite{Denbleyker:2010sv, Bazavov:2010xh, Denbleyker:2011aa}. This has been shown in the statistical spin models. This motivated us to calculate and analyze Fisher zeros of the lattice gauge theories with an IRFP and see how the RG flows behave. 

\section{Renormalization group flows}

RG flows are defined in some parameter space and the change of the parameters under RG transformation can be described by beta functions. Reliably calculating beta functions for systems near or inside the conformal window is very difficult \cite{DeGrand:2010ba}. In the following, we first  describe the RG flows qualitatively and then discuss how quantitative information can be learned from the Fisher zeros.

\begin{figure}
\begin{center}
\vspace{0.3in}
\includegraphics[width=0.45\textwidth]{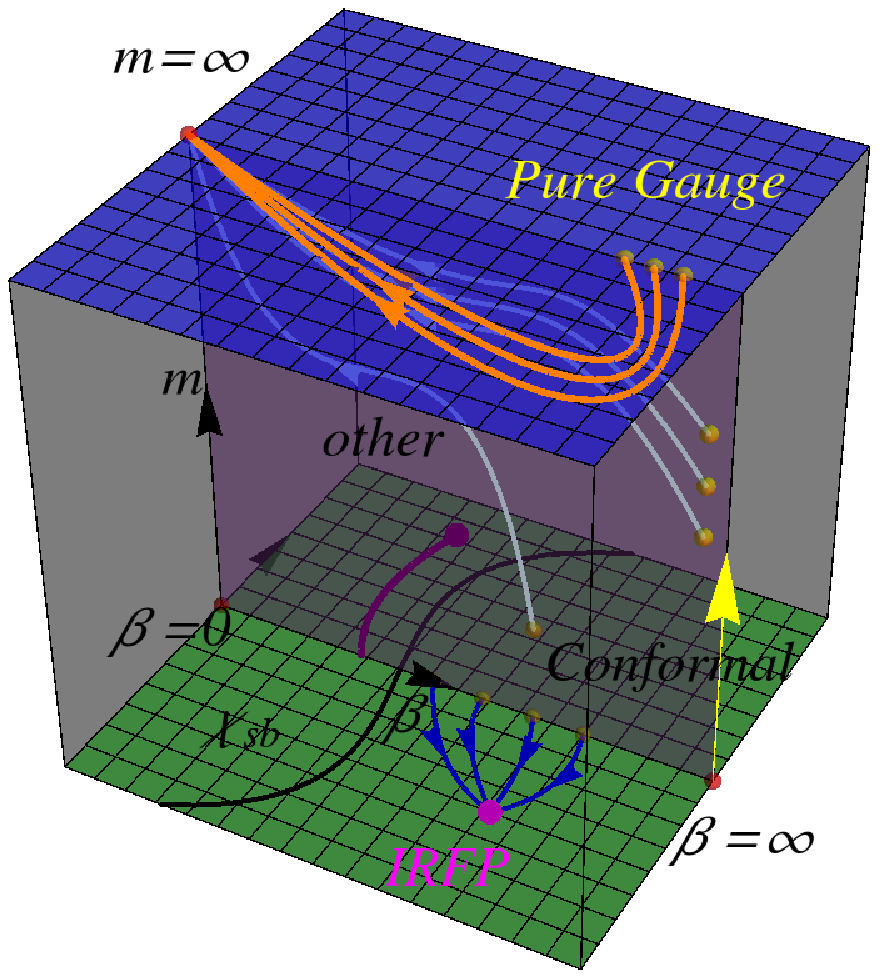}
\hspace{0.4in}
\includegraphics[width=0.45\textwidth]{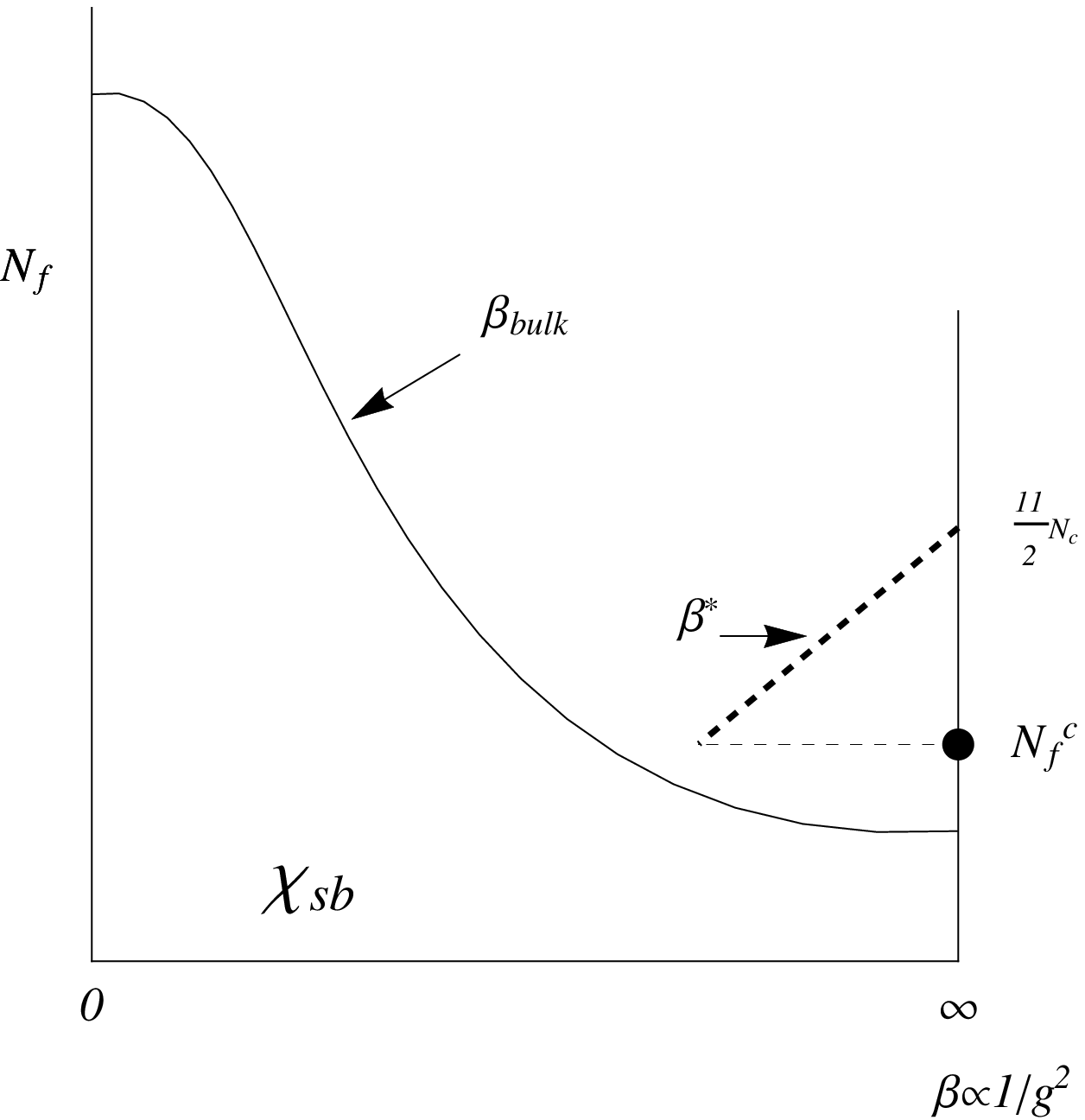}
\caption{Left panel: Schematic phase structure and flow diagram for some gauge theory with an IRFP. Right panel: Possible phase diagram in the $N_f$ vs. $\beta$ plane.}
\label{fig:RGFlows}
\vspace{-0.3in}
\end{center}
\end{figure}

The bare parameters are the parameters appearing in the simulated lattice action, such as $\beta$ and $m$ for  staggered fermions. The bare theory is a point in the $\beta$-mass plane. Effective parameters are parameters appearing after some coarse-graining procedure. For example, if we start with a pure gauge Wilson action, after one step Migdal-Kadanoff approximation other interaction terms will appear. One usually uses $\beta_\textnormal{Adjoint}$, $\beta_{3/2}$, $\dots$ to label them although the parametrization is not unique. 
In the following discussion, we will use $\beta$ to represent the bare coupling, $m$ the bare mass, and ``others'' all the other effective couplings. In Fig.~\ref{fig:RGFlows}, we represent  graphically the many dimensional space as three dimensional. 

For $SU(N_c)$ gauge theories with $N_f$ fermion flavors in some representation of the gauge group, whether there is an IRFP depends on $N_c$, $N_f$, and the representation. In the following, we will consider a generic theory with an IRFP. Our understanding of the phase structure and possible RG flows for such theories is summarized in Fig.~\ref{fig:RGFlows}.
The RG flows considered here always start from the ``bare'' plane, the $\beta$-mass plane, which is the purple vertical plane in Fig.~\ref{fig:RGFlows}. When the quarks are infinitely heavy ($m = \infty$), the theories are pure gauge. Therefore, the pure gauge plane is the upper blue plane in Fig.~\ref{fig:RGFlows}. In the dynamical simulations, ultimately one would like to extrapolate to $m = 0$ and the massless plane is the lower green plane in Fig.~\ref{fig:RGFlows}. If a theory has an IRFP it should be in this massless plane. 

In the vertical purple bare plane, the mass is the relevant parameter and the RG flows in the weak-coupling ($\beta = \infty$) limit will move vertically into the $m = \infty$ direction. If one starts from any other point with $m>0$ in the bare plane, the flows will all go to the strong-coupling ($\beta = 0$) limit $m = \infty$ point. These are shown as yellow and gray blue arrows in Fig.~\ref{fig:RGFlows}.

In the upper blue pure gauge plane, the flows will also go from the weak-coupling to the strong-coupling limit. As mentioned in the previous section, effective couplings will be generated during the RG transformation and the exact behavior of the RG flows will depend on the parametrization used. The orange flow lines describe some possible flows. 

In the lower green massless plane, chiral symmetry should be broken in the strong enough coupling region provided $N_f$ is not too large. Therefore, there is a chiral phase transition, which is usually signaled by a chiral condensate. When one is a little away from the massless plane, the chiral transition line will extend up to a surface (in the simplified picture) and in general one does not know the properties of this boundary region. The intersection of this surface with the bare plane 
is represented as a one dimensional darker purple line with an endpoint in Fig.~\ref{fig:RGFlows}.

If the theory has an IRFP, the properties of the theory are governed by the IRFP in the conformal phase. All the flows starting from a bare $\beta$ value in the conformal side of the massless plane will flow into the IRFP. In this sense, the bare coupling $\beta$ is an irrelevant parameter. 

This completes our discussion about the RG flows for a fixed $N_f$ inside the conformal window. When $N_f$ changes, it was shown \cite{deForcrand:2012vh} that the chiral symmetry may be restored for large $N_f$. Therefore, the chiral symmetry breaking phase transition line shown in the massless plane will disappear in the large $N_f$ limit. For $N_c = 3$ in the fundamental representation, the critical $N_f$ is around 52 staggered fermions. 
For a strong coupling discussion of chiral symmetry restoration see Ref. \cite{Tomboulis:2012nr}. 
This is shown in the right panel of Fig.~\ref{fig:RGFlows} in the $N_f$-$\beta$ plane. The boundary marked as $\beta_\text{bulk}$ is the separation line between vanishing and non-vanishing chiral condensate observed in the actual simulation and is of first order. The $N_f^c$ marked in the weak coupling  limit is the lower bond of the conformal window. We now proceed to explain the meaning of $\beta^*$.   

\begin{figure}[htp]
\begin{center}
\includegraphics[width=0.45\textwidth]{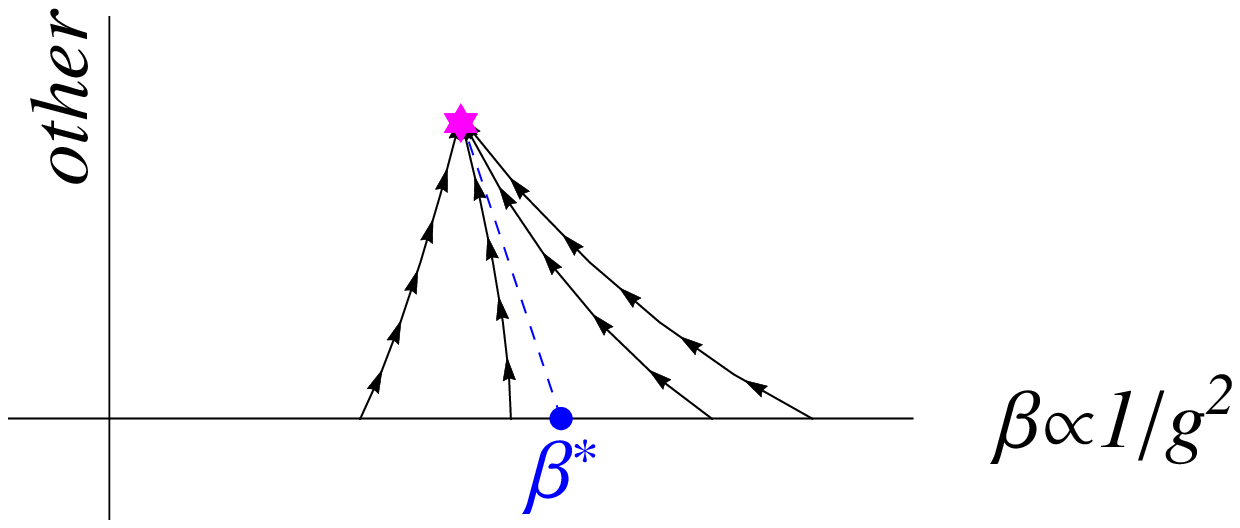}
\hspace{0.4in}
\includegraphics[width=0.45\textwidth]{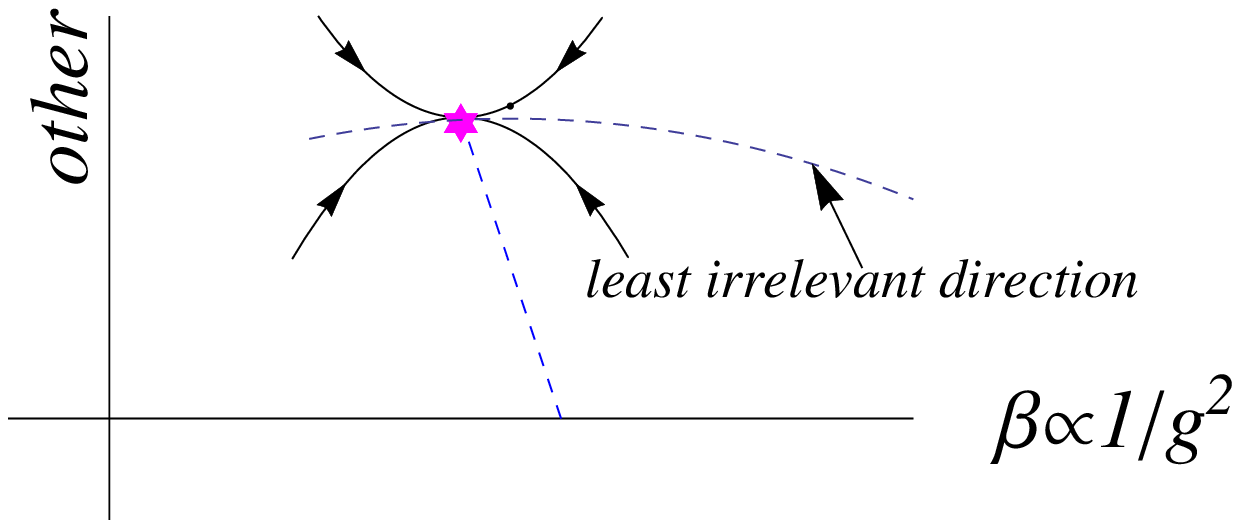}
\caption{\label{fig:flow2} Possible RG flows in the other-$\beta$ plane. Left panel: Systems inside but far away from the conformal window. Right panel: Systems inside and close to the edge of the conformal window.
}
\end{center}
\end{figure}

In the strictly massless case and for $N_f$ in the conformal window, there is an IRFP. If we assume that the IRFP is completely attractive in the massless plane, an example in terms of nonlinear scaling variables is shown below. Let us assume $u_{1, n}$, $u_{2, n}$ are two scaling variables after $n$ step RG transformation and $\lambda_1$, $\lambda_2$ are the corresponding eigenvalues. If
\begin{equation}
\begin{cases}
    u_{1, n+1}  =   \lambda_1~u_{1, n}\\
    u_{2, n+1}   =   \lambda_2~u_{2, n}
\end{cases}
\end{equation}
with $1 > \lambda_1 > \lambda_2 $, then $u_1$ corresponds to the least irrelevant direction. For the sake of illustration, we consider the case where 
$\lambda_2= \lambda_1^2$. We will have 
\begin{equation}
u_{2, n} = {\lambda_2}^n u_{2, 0} = {\lambda_1}^{2n} u_{2, 0} = {u^2_{1, n}} \frac{u_{2, 0}}{{u^2_{1, 0}}} \propto {u^2_{1, n}}.
\end{equation} 
This can be shown from  Fig.~\ref{fig:flow2}. If the least irrelevant direction has an intersection with the bare $\beta$ axis then the crossing point is $\beta^*$ and  one can use two-lattice matching or other techniques on either side of the fixed point to locate the IRFP. This is usually the case for a system with an IRFP but far away from the edge of the conformal window. 
An extreme case is shown in the right panel of Fig.~\ref{fig:flow2} and then the $\beta^*$ would be meaningless. 

The dashed $\beta^*$ line in the right panel of Fig.~\ref{fig:RGFlows} does not extend to the chirally broken phase because a QCD-like system can not be both conformal and chirally broken. But whether the $\beta^*$ touches the $\beta_{\text{bulk}}$ is not known (one should also keep in mind that $N_f$ is discrete). 
The $\beta_{\text{bulk}}$ boundary corresponds to the infinite volume limit case. However, in the actual simulation, the lattice size is finite. When the lattice size is ${L_x}^3 \times L_t$ and $L_x >> L_t$, the simulation is of finite temperature and the system may have a finite temperature transition. As $L_t$ increases, the finite temperature transition will turn into a bulk transition \cite{Deuzeman:2012ee}.

Based on the above discussion, one could ask the following questions:
What is the lowest value of $N_f$ for which the finite temperature transition turns into a bulk
transition?
Will the Fisher zeros pinch the real axis like $L^{-2}$ ($\nu$=1/2, mean field for a free scalar) instead of $L^{-4}$ near the endpoint (for $m = m_c$ )?
Is it possible to find a hint of the IR fixed point from the behavior of the zeros over a broader $\beta$ interval as a function of $m$?
Are the two questions related?
In the following, we will show some numerical results to try to give some hints on the above questions.

\section{Numerical results}

\begin{figure}
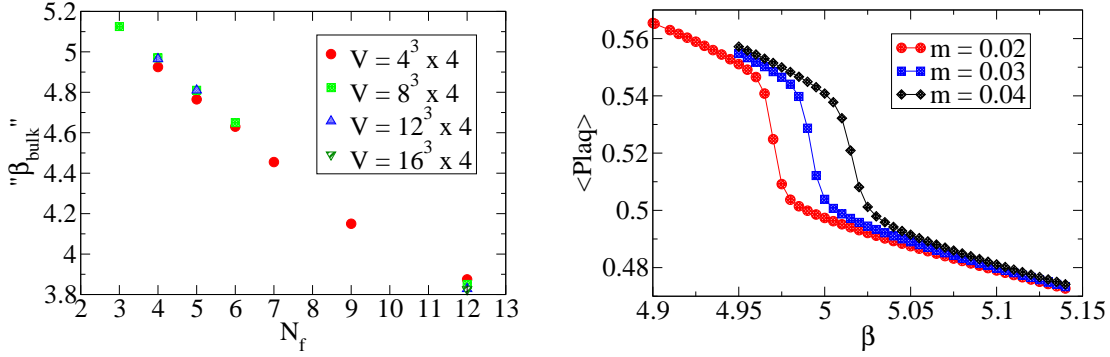

\begin{center}
\includegraphics[width=0.45\textwidth]{betac_nf2.eps}
\hspace{0.2in}
\includegraphics[width=0.47\textwidth]{mass.eps}
\caption{Left panel: ``$\beta_{bulk}$'' for various $N_f$ at fixed time dimension; the bare mass is fixed to be 0.02. Right panel: Average plaquette vs. $\beta$ at different masses.}
\label{fig:bulk_nf}
\end{center}
\end{figure}

We performed simulations with Wilson gauge action and naive staggered fermion action. The rational hybrid Monte Carlo technique allows us to simulate any numbers of fermion flavors at various bare parameters. 
We started the simulation with different $N_f$  but fixed bare mass 0.02. We simulated over a range of bare couplings for several different volumes. There is a clear discontinuity in both the average plaquette and chiral condensate. The location of the discontinuity moves to the weak-coupling region with increasing volume. For a fixed $L_t$, the location of the discontinuity moves from the weak-coupling region for small $N_f$ to the strong-coupling region for large $N_f$.
This can be seen from the left panel of Fig.~\ref{fig:bulk_nf}. The location also changes with changing $m$, which can be seen from the right panel of Fig.~\ref{fig:bulk_nf} for the four-flavors case.

\begin{figure}
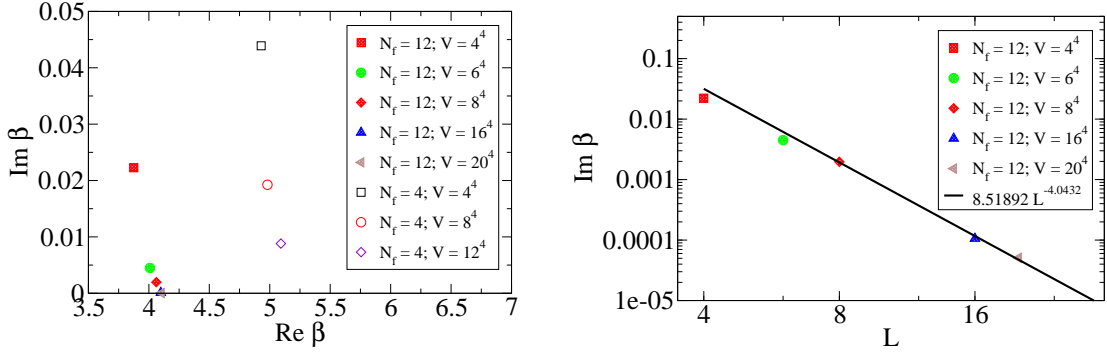

\includegraphics[width=0.45\textwidth]{zeros_nf4_nf12}
\hspace{0.2in}
\includegraphics[width=0.47\textwidth]{logbeta_logL_nf12}
\caption{Left Panel: Fisher zeros for $N_f=4$ and $N_f=12$. Right panel: Scaling of the lowest zeros with $L$ for $N_f=12$.}
\label{fig:zeros}
\end{figure}

From the simulations at different $\beta$, we can construct the density of states and thus calculate the Fisher zeros \cite{Denbleyker:2008ss, Denbleyker:2008zd}. The distribution of the zeros for $N_f = 4$ and $N_f = 12$ systems are clearly different. From the left panel of Fig.~\ref{fig:zeros}, we can see that the zeros for the $N_f = 12$ system has the tendency to pinch the real $\beta$ axis, which is a signal of a phase transition. Whereas the $N_f = 4$ system zeros are well above the real $\beta$ axis. 
The scaling of the zeros for $N_f = 12$ system is consistent with a first order phase transition.

\section{Conclusions}

In summary, we explain that RG flows can be understood from the Fisher zeros point of view; we also discuss in detail some possible RG flows in the reduced parameter space. For our preliminary numerical simulations, we could see  
a clear first order phase transition for $N_f = 12$ from the scaling of the zeros.
We are looking for the smallest $N_f$ for which the finite temperature transition turns into a bulk transition. It may be possible to make a connection between $\beta_{\text{bulk}}$ and $\beta^*$.
In principle, one could understand the mass dependence of the transition and calculate the mass anomalous dimension $\gamma_{m}$ from the Fisher zeros. As of now, the simulation is done for naive staggered fermions. 
Adding improvement terms to the fermion action may change the phase structure. 
We should mention that it is possible to perform the RG blocking via some newly developed tensor renormalization group methods \cite{Liu:2013nsa, Meurice:2013cla}.

This research was supported in part  by the Department of Energy under Award Numbers DE-SC0010114 and FG02-91ER40664. 
We used National Energy Research Scientific Computing Center, which is supported by the Office of Science of the U.S. Department of Energy under Contract No. DE-AC02-05CH11231. 
Y. L. is supported by the URA Visiting Scholars' program. 
Fermilab is operated by Fermi Research Alliance, LLC, under Contract No.~DE-AC02-07CH11359 with the United States Department of Energy. 
Y. M. did part of the work while at the workshop ``LGT in the LHC Era" in summer 2013 at the Aspen Center for Physics supported by NSF grant No 1066293.

\end{document}